\documentclass[aps,pre,superscriptaddress,twocolumn,showpacs]{revtex4-1}

\usepackage{graphicx}
\usepackage{CJK}
\usepackage{bm}

\usepackage{color}
\usepackage{graphicx}

\usepackage{hyperref}

\begin{document}


\title{Correlations Induced by Depressing Synapses in Critically Self-Organized Networks with Quenched Dynamics}



\author{Jo\~ao Guilherme Ferreira Campos}%
\email{joaogfc@gmail.com}
\affiliation{Departamento de F{\'\i}sica,
Universidade Federal de Pernambuco, 50670-901 Recife, PE, Brazil}%
\thanks{corresponding author}

\author{Ariadne de Andrade Costa}
\affiliation{Department of Psychological and Brain Sciences, Indiana University, 47405, Bloomington, IN, USA}
\affiliation{Instituto de Computa{\c c}\~ao,
Universidade Estadual de Campinas, 13083-852, Campinas, SP, Brazil}%

\author{Mauro Copelli}
\affiliation{Departamento de F{\'\i}sica,
Universidade Federal de Pernambuco, 50670-901 Recife, PE, Brazil}%

\author{Osame Kinouchi}
\affiliation{Departamento de F{\'\i}sica, FFCLRP,
Universidade de S\~ao Paulo, 14040-901, Ribeir\~ao Preto, SP, Brazil}%




\begin{abstract}
In a recent work, mean-field analysis and computer simulations were employed to analyze critical 
self-organization in networks of excitable cellular automata where randomly chosen synapses in the network
were depressed after each spike (the so-called annealed dynamics). 
Calculations agree with simulations of the annealed version, 
showing that the nominal \textit{branching ratio\/} $\sigma$ converges to unity 
in the thermodynamic limit, as expected of a self-organized critical system. However, 
the question remains whether the same results apply to the biological case
where only the synapses of firing neurons are depressed (the so-called quenched dynamics). 
We show that simulations of the quenched model yield significant deviations from $\sigma=1$ due 
to spatial correlations. However, the model is shown to be critical, as the largest eigenvalue 
of the synaptic matrix approaches unity in the thermodynamic limit, that is, $\lambda_c = 1$ . 
We also study the finite size effects near the critical state as a function of the parameters of the synaptic dynamics.
\end{abstract}

\pacs{05.65.+b, 05.70.Ln, 07.05.Mh}

\maketitle

\section{Introduction} 

The first empirical evidence of criticality in the brain was given by Beggs and Plenz, 
who reported that in vitro rat cortex slices exhibit neuronal avalanches with power law distribution with 
exponents -3/2 and -2 for avalanche size and duration, respectively~\cite{Beggs03}. This was regarded as 
evidence that the brain as a dynamical system fluctuates around a critical point, presumably similar to a 
critical branching process with the branching parameter close to unity. Either on theoretical or experimental 
grounds, this property has been shown to optimize computational capabilities~\cite{Bertschinger04}, 
information transmission~\cite{Beggs03,Shew11}, sensitivity to stimuli 
and enlargement of dynamic range~\cite{Kinouchi06a,Shew09,
Gautam2015,GirardiShappo2016}, among others, as recently reviewed in Refs.~\cite{Chialvo10,Shew13}.

In order to explain the self-organization around the critical point, Arcangelis \emph{et al.}~\cite{deArcangelis06} 
introduced a model with synaptic depression and synaptic recovery (see also~\cite{deArcangelis12,deArcangelis16}). 
They obtained Self-Organized Criticality (SOC) and other very interesting results but their synaptic depression mechanism has the 
undesirable feature of depending on non-local information. In contrast,
Levina, Herrmann and Geisel (LHG) proposed a local model which consists of a fully 
connected network of integrate and fire neurons, such that when a neuron fires, the strength of its 
output links (synapses) is reduced by a fraction~\cite{Levina07}. This fast dissipation mechanism has been associated with 
short-term synaptic depression due to temporary neurotransmitter vesicle depletion. It is countered by a  
recovery mechanism at a different time scale, by which synaptic neurotransmitters are slowly replenished 
when the synapse is idle. LHG claimed their model exhibit SOC, 
based on the emergence of power-law distributions of avalanche sizes. 

On a pair of review papers, however, Bonachela \textit{et al.} showed that for a system to exhibit 
SOC its bulk dynamics must be conservative at least on average~\cite{Bonachela09, Bonachela10, 
Moosavi&Montakhab}. They showed that systems with dissipative and loading mechanisms such as the 
LHG model would hover around the critical point with nonvanishing fluctuations even in the
thermodynamic limit. In that sense, the behavior of the LHG model 
would not be classified as SOC, but rather was called Self Organized quasi-Criticality (SOqC). 
Indeed, the LHG model seems to pertain to the Dynamical Percolation universality class~\cite{Bonachela10} and
not to the Directed Percolation class as is usual in \emph{bona fide} SOC models.   

In a recent work, we analyzed a random neighbor network of excitable cellular automata with 
dynamical synapses~\cite{Costa15}. It was inspired by the  LHG model, but with three different 
ingredients: finite connectivity (with $K$ outgoing synapses in a random graph), discrete state units 
and multiplicative probabilistic synapses. In the so-called annealed version of the synaptic dynamics, 
when some neuron spikes, another neuron is chosen randomly in the network and its synapses are depressed.
That is, there is no correlation between the firing locus and the synapses that are depressed.  This artificial annealed
procedure has been introduced because mean field calculations describes perfectly this case, see Ref.~\cite{Costa15}.

This model was shown to behave very differently from the 
LHG model, in that not only the stationary temporal average branching ratio 
$\left\langle \sigma(t) \right\rangle$ converged to unity but, 
perhaps most importantly, the fluctuations around the criticality condition $\sigma_c =1$ vanished 
in the thermodynamic limit $N \rightarrow \infty$. 
Also, the associated phase transition is standard: continuous, with one absorbing state, 
and in the Directed Percolation universality class, as shown by simulations and  mean-field results.

Despite the agreement between the mean-field calculations and the numerical simulations, 
a major drawback of the annealed model is its lack of biological plausibility. Here we investigate 
in detail the quenched version of the model, in which, when a presynaptic neuron fires, only its 
outgoing synapses are depressed. In particular, we focus on whether the quenched model behaves 
similarly to the annealed model as far as SOC is concerned. 

The rest of the paper is structured as follows. In section 2, we revisit the model, and in particular 
the differences between annealed and quenched synaptic dynamics. In section 3, we present our simulation 
results and discussions. Concluding remarks appear in section 4.


\section{The Model} 

Our model builds upon a random-neighbor network of excitable automata neurons [quiescent ($S_j=0$), to 
firing ($S_j = 1$), to refractory ($S_j = 2, 3,\dots, n-1$), to quiescent ($S_j = 0$, $j = 1,\dots, N$)] 
used previously~\cite{Kinouchi06a,Costa15}. In this version, $N$ sites with states $S_j (t)$ 
have each exactly $K_{j}^{out}=K$ outlinks randomly chosen to postsynaptic neurons $S_i (t)$. 
With this construction, each neuron has $K_{j}^{in}$ binomially distributed incoming links with average $K$. 
The adjacency matrix $A_{ij} \in [0,1]$ is fixed from the start and never changes, defining the neighborhood topology. 
This is not exactly a canonical Erd{\H o}s-R\'enyi (ER) network, which also has binomially distributed outlinks, but it is
very close, so that we say that our networks have an ER-like topology.

Only for pairs that have $A_{ij}=1$ we have probabilistic synapses $0 < P_{ij} < 1$. This means that,
if presynaptic neuron $j$ fires, then, at the next time step, postsynaptic neuron $i$
fires with probability $P_{ij}$  (updates are done in parallel). Since neuron $i$ has an average of $K$ presynaptic neighbors, there occurs an average of 
$K$ independent attempts. 
After a site spikes, it deterministically becomes refractory for $n- 2$ time steps [$S_i(t + 2) = 2$, $S_i(t + 3) = 3$, 
$\dots$, $S_i (t+ n- 1) = n-1$] and then returns to quiescence $S_i=0$ (for details, see Ref.~\cite{Costa15}).
When the network falls in the absorbing state (no firings), a randomly chosen site is
forced to fire so that the network activity returns.

The dynamics on the synapses can be of three kinds:
\begin{itemize}
\item The fixed case: $P_{ij}$ are fixed, and never changes, as studied in Ref.~\cite{Kinouchi06a};
\item The quenched case: $P_{ij}(t)$ vary in time following a local rule that preserves spatial correlations;
\item The annealed case: $P_{ij}(t)$ follow a global rule that does not preserve spatial correlations, as in Ref.~\cite{Costa15}.
\end{itemize}

For the  quenched case, the synapses obey the following equation:
\begin{equation}
\label{Pij}
    P_{ij}(t+1) = P_{ij}(t) + \frac{\epsilon}{KN^a} (A - P_{ij}(t)) - u P_{ij}(t) \delta(S_j(t)-1),
\end{equation}
where $\epsilon$ is the coefficient of synaptic recovery, $A$ is the asymptotic synaptic value and $u$ is 
the fraction of coupling strength that is lost whenever a neuron fires, related to short time depletion
of synaptic neurotransmissor vesicles~\cite{Levina07,Costa15}. 
The Kronecker $\delta(x)$ is one for $x=0$ and zero otherwise. 
This synaptic dynamics means that, whenever the neuron $j$ fires, all its outgoing 
synapses are reduced to basically to ($1-u$) of their original value, since the second term is small. 
The exponent $a$ enable us to explore how the model behaves with different scalings for $N$ 
in  the synaptic recovery dynamics.

For annealed dynamics, instead of depressing the $K$ outgoing synapses of the firing  
neuron $j$, either $K$ randomly chosen synapses are depressed or a random 
neuron is chosen and its $K$ outgoing synapses are depressed. 
The purpose of the annealed dynamics is to destroy correlations between the $P_{ij}$, so one can use mean-field 
analysis to get a better insight of the problem, as done previously~\cite{Costa15}. 
We have tested both types of annealing,
and they  work equally well in destroying correlations, but the latter is  computationally more efficient. 

Of course, from a realistic or biological point of view, the annealed case does not make sense, only the quenched dynamics.
In any case, we must emphasize that both dynamics for $P_{ij}(t)$ never change the structure of the neighborhood
given by $A_{ij}$, that is, both annealed and quenched dynamics take place on topologically ER-like networks.

The system is set in the slow driving limit. The initial condition is $S_i(0) = 0$ for all $i\neq k$ 
and $S_k(0)=1$, i.~e., we start an avalanche at site $k$. Whenever the system returns to quiescence
(that is, the absorbing state $S_i(t) = 0$, $\forall i$), we start another avalanche
by choosing a random neuron $m$, and setting $S_m(t+1)=1$. In each time step we apply the
synaptic dynamics Eq.~(\ref{Pij}) or its annealed version.

The initial conditions for the synapses are defined by choosing the initial average synaptic value $\sigma_0/K$ 
and uniformly drawing random values to $P_{ij}(0)$ in the interval  $\left[0, 2 \sigma_0/K\right]$. 
At each time step we compute a local branching ratio $\sigma_j^{out}(t) = \sum_{i=1}^{K} P_{ij}(t)$ 
and a global branching ratio $\sigma(t) = \frac{1}{N}\sum_{j=1}^N \sigma_j^{out}(t)$. After a transient,
$\sigma(t)$ fluctuates around some average value $\sigma^*$ with standard deviation $\Delta\sigma^*$. 

As shown previously~\cite{Costa15}, the mean-field analysis predicts that
\begin{equation}
    \sigma^*  \simeq   1+ \frac{(AK-1)}{1+x},
    \label{sigmaMF}
\end{equation}
\noindent where $x \equiv uKN^a/[(n-1)\epsilon]$. 

This result only holds for  $A > 1/K$ (notice that perfect criticality can be achieved with 
the choice $A = 1/K$, but this is a fine tuning for parameter  $A$  that should not be used 
in the SOC context).  In the limit $x \gg 1$, we get:
\begin{equation}
\label{Omega-N}
    \sigma^* \simeq    1   +   \frac{\Omega}{N^a} \:,
\end{equation}
where $\Omega \equiv (AK-1)(n-1)\epsilon / (uK)$. That is, the mean-field analysis predicts 
that, for $a>0$ and large $N$, $\sigma^*$ differs from $\sigma_c=1$ by a factor of order 
$1 / N^a$, therefore $\sigma^* \rightarrow 1$ in the infinite-size limit. 
The case $a=0$ will be discussed separately.

We define the Perron-Frobenius (largest) eigenvalue of the connectivity matrix $P_{ij}(t)$ as 
$\lambda(t)$, which in the stationary state fluctuates around the mean value $\lambda^*$ with 
standard deviation $\Delta\lambda^*$. 
As shown previously by Larremore \emph{et al.}~\cite{Larremore11a}, 
the phase transition between an absorbing and an active phases occurs generally at $\lambda_c=1$, not necessarily at $\sigma_c = 1$. 
However, in~\cite{Larremore11a} these results are derived for static networks, with fixed $P_{ij}$, and spatial correlations between nodes are imposed \emph{a priori}.

Here, our problem is different: we start from an ER-like network without correlations between the sites.
For the fixed synapses without correlations case we have $\lambda = \sigma$~\cite{Kinouchi06a,Larremore11a}.
However, here the $P_{ij}(t)$ are not fixed but evolve. 
Our central questions are: 
\begin{itemize}
\item Does the quenched synaptic dynamics produce $\lambda^* \simeq \lambda_c = 1$ or $\sigma^* \simeq \sigma_c = 1 $? 
\item How do the correlations between synapses, necessary to produce $\lambda \neq \sigma$, arise? 
\item How do the fluctuations around $\lambda^*$ behave as a function of $N$? 
\end{itemize}

In the next section, we analyze the behavior of $\lambda^*$ and $\sigma^*$, as well as their respective 
standard deviations, varying our model parameters $\epsilon$, $u$, $A$, $K$ and the network size $N$. 
We concentrate on the case $a=1$ since all previous literature 
examined  this scaling~\cite{Levina07, Bonachela10,Costa15}, but we also discuss briefly other values for the exponent $a$.


\section{Simulation Results}

\begin{figure}[!h]
\centering
\includegraphics[width=1.0\columnwidth]{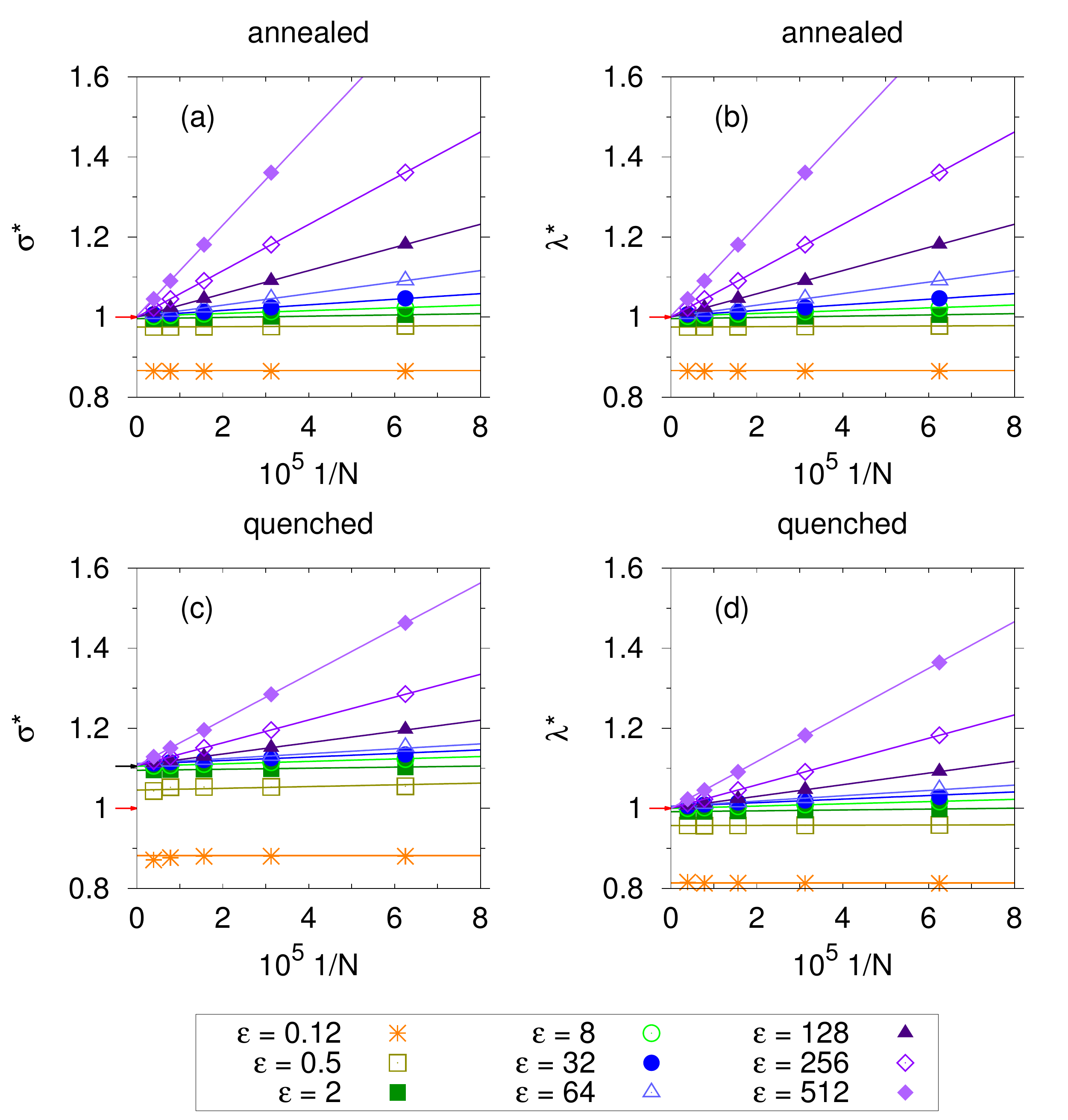}
\caption{$\sigma^*$ and $\lambda^*$ versus $1/N$ for several values of $\epsilon$, 
for quenched and annealed synaptic dynamics. The error bars do not appear at this scale. 
The lines are curves of the type $f(N) = \alpha + \beta / N$ that best fit the data, 
except for (a) where the curves are given by Eq.~\ref{Omega-N} for 
$\epsilon \geq 8$. Parameters: $n=3$, $K=10$, $A=1.0$, $u=0.1$, $a=1$. 
The small arrows point to the value 1 and, in (c), also to 1.105.}
\label{Lambda-Sigma-vs-N-e-FIT}
\end{figure}

To simplify our simulations we fix $K=10$, meaning that the number of outgoing synapses is much 
smaller than the number of neurons. If we fix $\epsilon$ and plot $\sigma^*$ and $\lambda^*$ while 
we vary $N$ (with exponent $a=1$), we obtain Fig.~\ref{Lambda-Sigma-vs-N-e-FIT}. Observe in 
Fig.~\ref{Lambda-Sigma-vs-N-e-FIT}a that, for annealed dynamics, when $\epsilon$ is small (say, $0.12$ or $0.5$), 
then $\sigma^*$  is smaller than 1 and independent of $N$. 
We call this regime the subcritical one, that is,
there is a large volume in parameter space $(A,\epsilon,u)$ where no self-organization to criticality exists.
However,  as $\epsilon$ increases, $\sigma^*$
also increases until it starts behaving like Eq.~\ref{Omega-N} for $\epsilon \gtrsim 8$.
Indeed, Eq.~\ref{Omega-N} has been derived, and is only valid, above the critical point~\cite{Costa15}.
For these values of $\epsilon$, we also plot in Fig.~\ref{Lambda-Sigma-vs-N-e-FIT}a the 
curves predicted by Eq.~\ref{sigmaMF}.

\begin{figure}[!h]
\centering
\includegraphics[width=1.0\columnwidth]{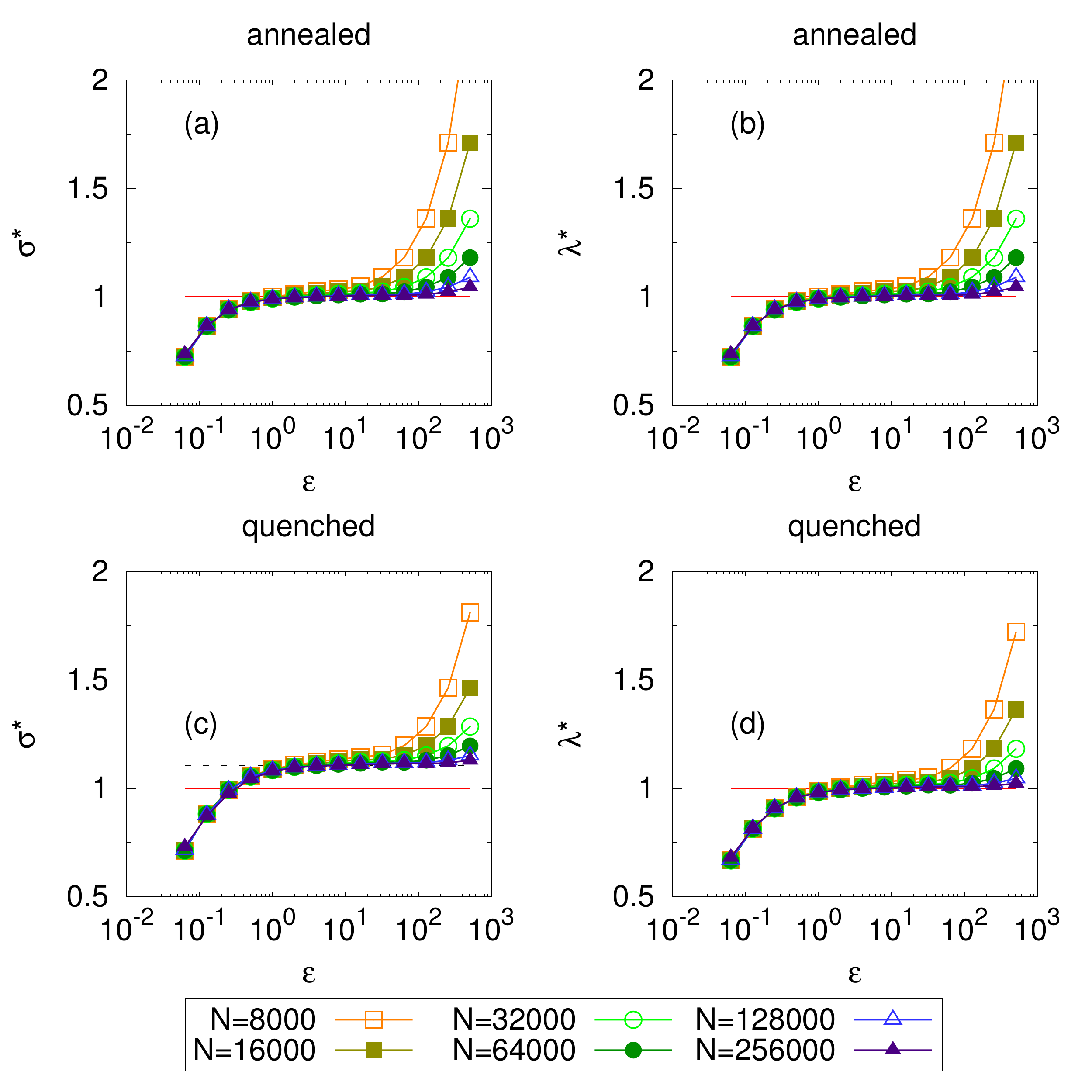}
\caption{$\sigma^*$ and $\lambda^*$ versus $\epsilon$ for several values of $N$ for 
quenched and annealed dynamics. 
The error bars do not appear in this scale. The horizontal line is the critical value $1$ and the dashed line is $1.105$. The other lines are guides to the eye. 
Parameters: $n=3$, $K=10$, $A=1.0$, $u=0.1$, $a=1$.}
\label{Sigma-Lambda-vs-Epsilon}
\end{figure}

For quenched dynamics, the behavior is quite similar, as shown in Fig.~\ref{Lambda-Sigma-vs-N-e-FIT}c. 
However, as we increase $\epsilon$, $\sigma^*$ behaves as: 
\begin{equation}
    \sigma^*    \simeq   1.105 +  \frac{\Omega_q}{N},
\end{equation}
\noindent for some constant $\Omega_q$, instead of following Eq.~\ref{Omega-N}. 
That indicates that mean-field theory does not describe well the quenched case. 
Furthermore, if we fix $N$ and vary $\epsilon$, as in Fig.~\ref{Sigma-Lambda-vs-Epsilon},
 we see that for the annealed case (Fig. \ref{Sigma-Lambda-vs-Epsilon}a) we obtain $\sigma^*\approx 1$ 
for a wide range of the parameter values (a plateau) that gets larger as $N$ increases. For 
quenched dynamics (Fig.~\ref{Sigma-Lambda-vs-Epsilon}c), we see the same behavior with 
$\sigma^*\approx1.105$ instead of $1$. This strange behavior for quenched 
dynamics does not appear in the plots for $\lambda^*$ shown in Fig.~\ref{Lambda-Sigma-vs-N-e-FIT}d 
and \ref{Sigma-Lambda-vs-Epsilon}d. 

\subsection{The relation between $\lambda$ and $\sigma$}
All this occurs because $\sigma^*$ is the wrong ``control'' parameter: $\lambda$ is the correct predictor for criticality, 
as shown by Restrepo \emph{et al.} (2007)~\cite{Restrepo07} and Larremore \emph{et al.}~\cite{Larremore11a}.
Indeed, these authors derived a good approximation for networks with homogeneous degree, like ours, which states that:
\begin{eqnarray}
\lambda = \eta \sigma \:, \\
\eta = \frac{\left\langle \sigma_i^{in} \sigma_i^{out} \right\rangle }{\sigma^2} \:,
\end{eqnarray}
where $\eta$ was called the correlation coefficient and $\sigma_i^{in} = \sum_j P_{ij}$ is  the sum of incoming links.
The average $\left\langle \ldots \right\rangle $ is over the sites $i$.
So, the $\eta$ coefficient measures if incoming and outcoming synapses are correlated:
$\eta = 1, \lambda = \sigma$ for uncorrelated synapses (since $\sigma = \left\langle \sigma_i^{in} \right\rangle =
 \left\langle \sigma_i^{out} \right\rangle$), $\eta > 1, \lambda > \sigma $ for correlated synapses and 
 $\eta < 1, \lambda < \sigma$ for anti-correlated synapses.

\begin{figure}[!h]
\centering
\includegraphics[width=0.68\columnwidth]{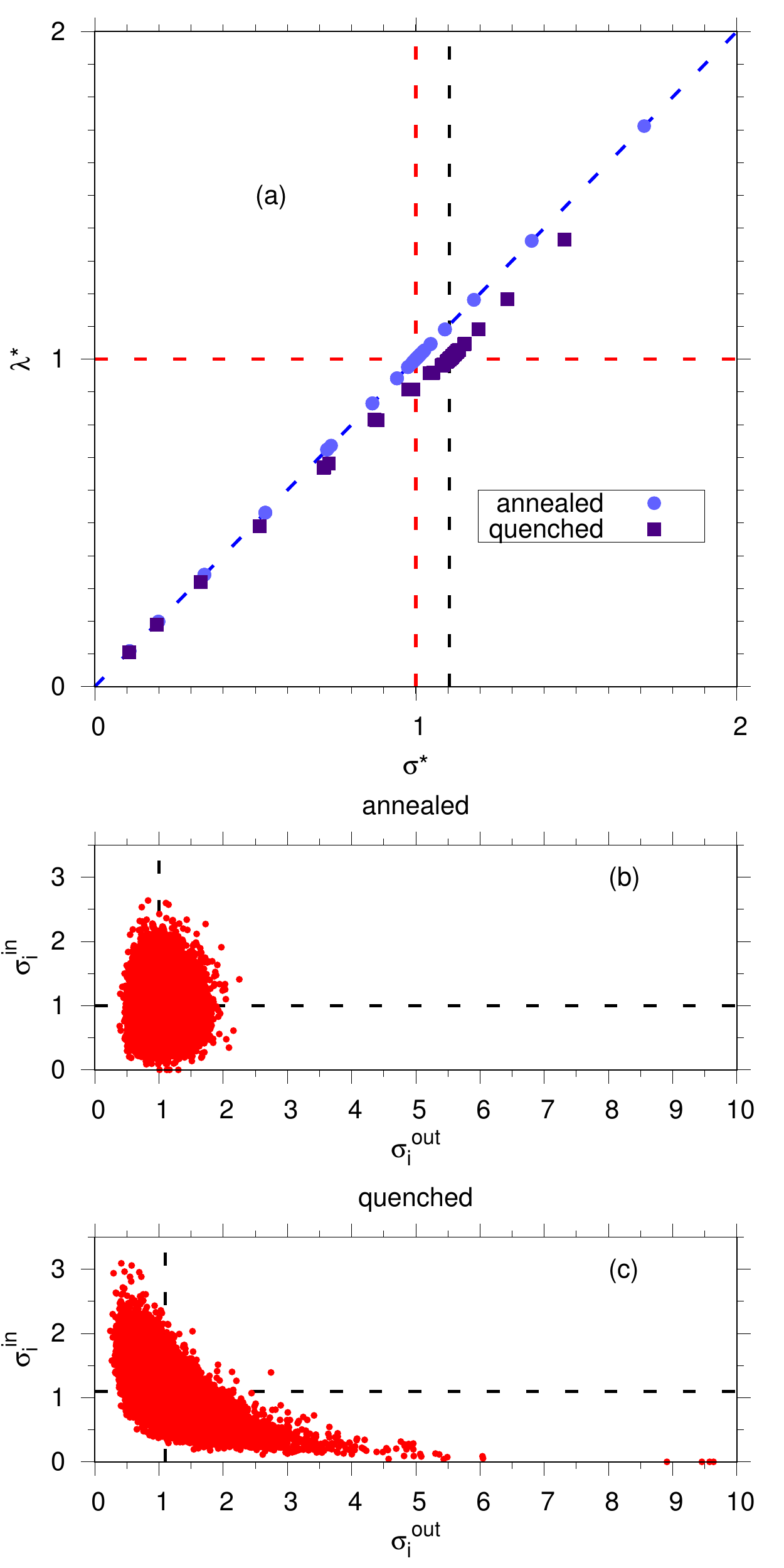}
\caption{(a) $\lambda^*$ versus $\sigma^*$ for several values of $N$ and $\epsilon$ 
for the quenched and annealed cases. The horizontal dashed line is $\lambda^*=1$ and the 
vertical dashed lines are $\sigma^*=1$ and $\sigma^*=1.105$.  The diagonal dashed line is the identity function 
$\lambda^*=\sigma^*$. Other parameters: $n=3$, $K=10$, $A=1.0$, $u=0.1$, $a=1$. 
(b) and (c) $\sigma_i^{in}$ versus $\sigma_i^{out}$ for annealed and quenched dynamics, 
respectively, in the stationary regime, with $N=32000$ and $\epsilon=2$. The dashed lines are the curves $\sigma_i^{in},\sigma_i^{out}=1.0$ in the annealed case and $\sigma_i^{in},\sigma_i^{out}=1.105$ 
in the quenched case. 
Other parameters are the same as before. The Spearman correlation
coefficient for the annealed case (b) is $-0.002$ (no correlation), and for the quenched 
case (c) is $-0.696$ (strong negative correlation).}
\label{scatterplot}
\end{figure}

If we plot $\lambda^*$ versus $\sigma^*$ for all the simulations shown in 
Figs.~\ref{Lambda-Sigma-vs-N-e-FIT} and \ref{Sigma-Lambda-vs-Epsilon} we get Fig.~\ref{scatterplot}a. 
Notice that for quenched dynamics we obtain $\sigma^*=1.105$ exactly when $\lambda^*=1$ 
(Fig.~\ref{scatterplot}a), meaning that $\lambda^*$ indeed approaches a critical 
value  (see Fig.~\ref{Lambda-Sigma-vs-N-e-FIT} and \ref{Sigma-Lambda-vs-Epsilon}). 

In the annealed case, the points lie exactly in the identity curve (Fig.~\ref{scatterplot}a). 
This is consistent with the above result that the equality $\sigma=\lambda$ holds when $\eta = 1$ and there are no correlations between the $P_{ij}$ \cite{Restrepo07,Larremore11a}. 
It is also consistent with the idea that random networks with uncorrelated weights are analogous to mean-field standard branching processes, where $\sigma$ is the correct control parameter~\cite{Harris02}. 

For the quenched case, we have $\lambda^* < \sigma^*$ (Fig.~\ref{scatterplot}a), suggesting anti-correlation $\eta < 1$. 
The question still remains as to what these correlations are and how they arise. 
In Fig.~\ref{scatterplot}b, we plot $\sigma_i(t)^{in}$ versus $\sigma_i^{out}(t)$ 
in the stationary regime for  networks with annealed dynamics. 
The two quantities are uncorrelated (Spearman correlation coefficient
$-0.002$), compatible with $\eta = 1$ and $\lambda = \sigma$.
In Fig.~\ref{scatterplot}c we show the same result for quenched dynamics. 
Now the two quantities are negatively correlated (Spearman coefficient $-0.696$)
and, consistently, $\lambda < \sigma$. The negative 
correlation has an intuitive explanation: if a given site has a high (low) local 
converging ratio $\sigma_i^{in}$, it will spike more (less) often, depressing more (less) its synapses, 
which implies a lower (higher) probability $\sigma_i^{out}$ of exciting its neighbors. 

We would like to emphasize that the difference between $\sigma$ and $\lambda$ is not due to changes in the topology defined by $A_{ij}$. We are working with weighted ER-like networks and the dynamics is on the real-valued weights $P_{ij}(t)$, not on $A_{ij}$. It is not a change in the ER topology that produces $\lambda < \sigma$ but, for a node $j$, correlations between the incoming $P_{ji}$ and the outgoing $P_{kj}$.

\subsection{The limit $N \rightarrow \infty$} 
\begin{figure}[!h]
\centering
\includegraphics[width=1.0\columnwidth]{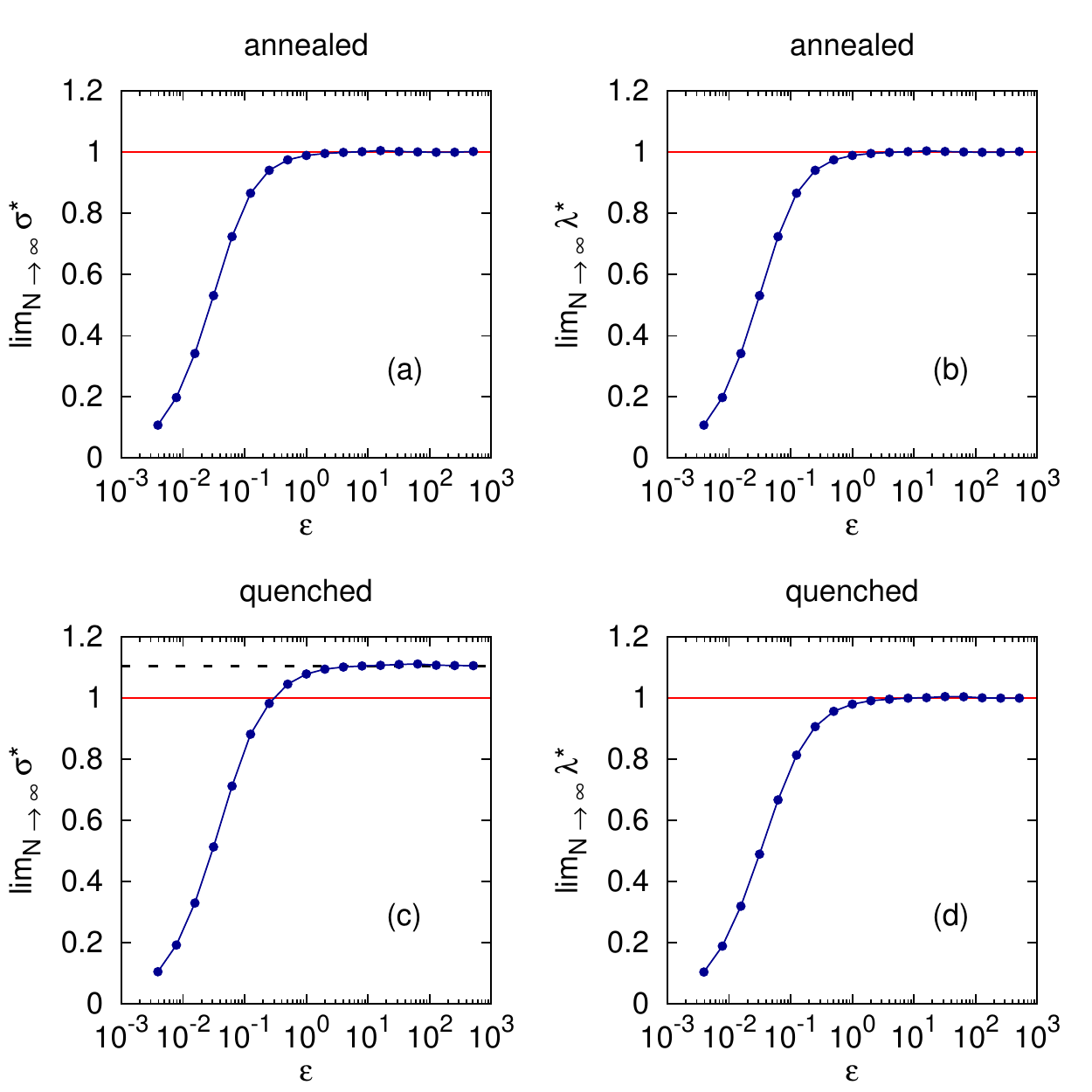}
\caption{The limit of $\sigma^*$ and $\lambda^*$ when $N\to\infty$ versus the 
parameter $\epsilon$ for quenched and annealed dynamics.
The horizontal line is the critical value $1$ and the dashed line is $1.105$. 
Curves are guides to the eye. 
Parameters: $n=3$, $K=10$, $A=1.0$, $u=0.1$ and $a=1$.}
\label{lim-N-infty-Sigma-Lambda-vs-Epsilon}
\end{figure}

In Fig.~\ref{Lambda-Sigma-vs-N-e-FIT} we see that all curves are straight lines. 
Thus, aiming at understanding the behavior of the networks as $N\to\infty$, we fit 
the data with curves of the type $f_{\sigma}(N)=\alpha_{\sigma}+\frac{\beta_{\sigma}}{N}$ 
and $f_{\lambda}(N)=\alpha_{\lambda}+\frac{\beta_{\lambda}}{N}$ for $\sigma^*$ and 
$\lambda^*$, respectively, as shown in Fig.~\ref{Lambda-Sigma-vs-N-e-FIT} 
(except for $\sigma^*$ with $\epsilon\geq 8$ in the annealed case, where we show 
the theoretical curve). We also observe that $\lim_{N\to\infty}\sigma^* = \alpha_{\sigma}$ and 
$\lim_{N\to\infty}\lambda^* = \alpha_{\lambda}$ and plot them as a function of $\epsilon$. 
The result is shown in Fig.~\ref{lim-N-infty-Sigma-Lambda-vs-Epsilon}. 

For networks with annealed dynamics and $\epsilon\gtrsim 4$, we obtain $\lim_{N\to\infty}\sigma^*=1$, 
within errors (Fig.~\ref{lim-N-infty-Sigma-Lambda-vs-Epsilon}a). 
For networks with quenched dynamics and $\epsilon\gtrsim 8$, we obtain $\lim_{N\to\infty}\sigma^*=1.105$, 
within errors (Fig.~\ref{lim-N-infty-Sigma-Lambda-vs-Epsilon}c). 
On the other hand, $\lim_{N\to\infty}\lambda^*=1$, within errors, for $\epsilon\gtrsim 4$ 
(annealed, Fig.~\ref{lim-N-infty-Sigma-Lambda-vs-Epsilon}b) and $\epsilon\gtrsim 8$ 
(quenched, Fig.~\ref{lim-N-infty-Sigma-Lambda-vs-Epsilon}d). 
That is, networks with quenched (annealed) dynamics are either subcritical, if $\epsilon\lesssim 8$ 
($\epsilon\lesssim 4$), or critical, if $\epsilon\gtrsim 8$ ($\epsilon\gtrsim 4$), 
in the infinite size limit, but never supercritical. 
This confirms that there is a semi-infinite volume in the $(\epsilon, A, u)$ parameter space 
with critical $\alpha_\lambda$. This is one of the most important results of the paper.

Since $\lambda \simeq \sigma$ for annealed dynamics, and since the behavior of $\lambda$ 
is qualitatively the same for both annealed and quenched dynamics, from now on we focus 
on quenched synaptic dynamics only. 

\subsection{Dependence on model parameters}
\begin{figure}[!h]
\centering
\includegraphics[width=1.0\columnwidth]{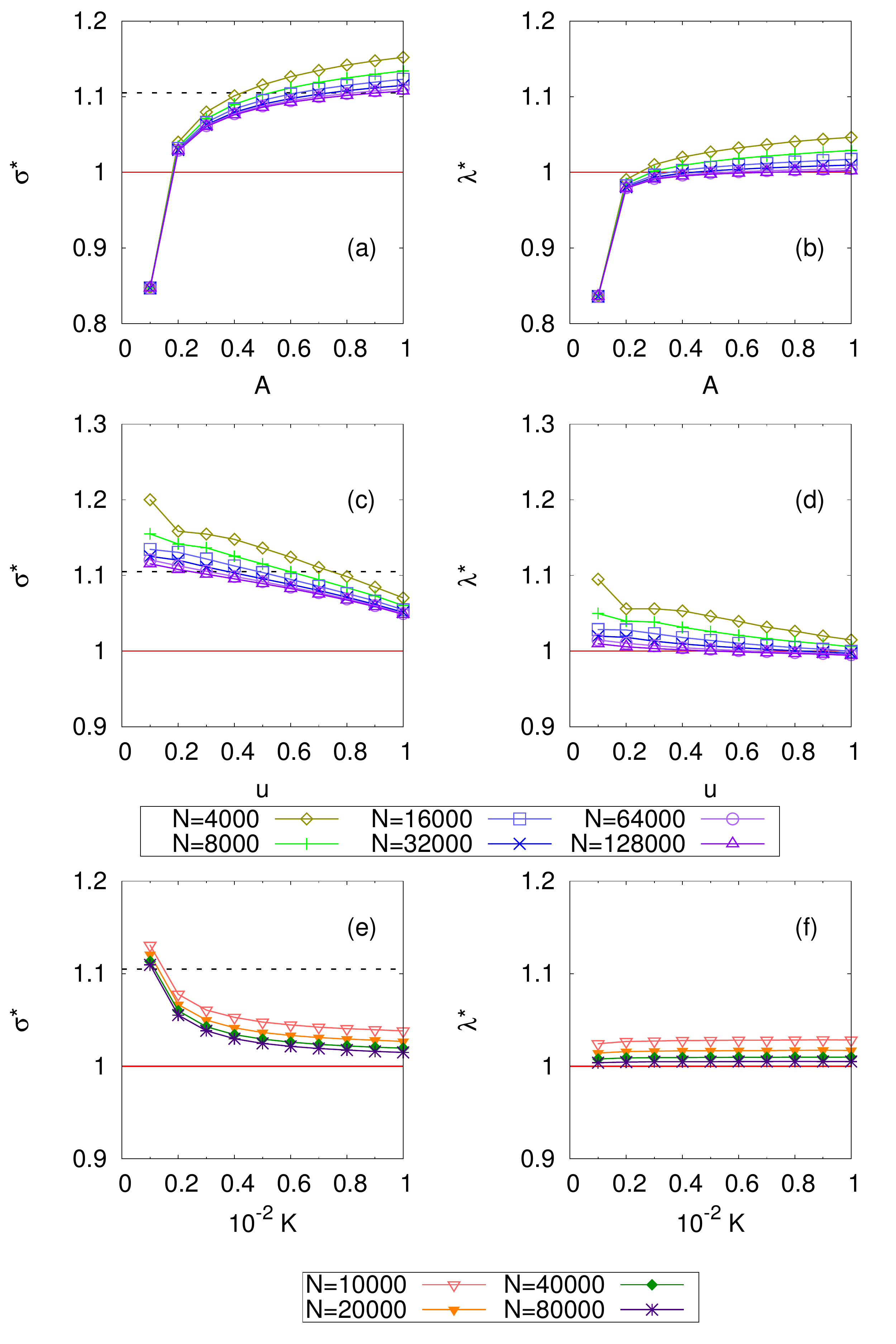}
\caption{(a) and (b) $\sigma^*$ and $\lambda^*$, respectively, versus $A$ as $N$ increases. 
Parameters: $\epsilon=8$, $n=3$, $K=10$, $u=0.1$, $a=1$. (c) and (d) 
$\sigma^*$ and $\lambda^*$, respectively, versus $u$ as $N$ increases. 
Parameters: $\epsilon=32$, $n=3$, $K=10$, $A=1.0$, $a=1$. (e) and (f) $\sigma^*$ and 
$\lambda^*$, respectively, versus $K$ as $N$ increases. 
Parameters: $\epsilon=8$, $n=3$, $A=1.0$, $u=0.1$, $a=1$. 
The error bars do not appear in the scales. 
The horizontal line is the critical value $1$ and the dashed line is $1.105$. 
The other lines are guides to the eye. 
The figure refers to quenched synaptic dynamics.}
\label{Sigma-Lambda-vs-uKA}
\end{figure}

\paragraph{Dependence on $A$:}
The influence of parameter $A$ on the values of $\sigma^*$ and $\lambda^*$, for different network sizes, is shown in Fig.~\ref{Sigma-Lambda-vs-uKA}a,b. 
The system is subcritical only for small values of $A$. 
This occurs because the recovery term $(A - P_{ij})$ implies that $P_{ij} \rightarrow A$ in the limit of no activity. 
Since we need $P_{ij} \gtrsim 1/K$ to 
achieve criticality, the condition  $A > 1/K = 0.1$ must be satisfied.
Lower values of $A$ take the network to subcritical states regardless of $\epsilon$ and $u$. 
Nevertheless, for $A > 1/K$, the system becomes critical as its size is increased.

\paragraph{Dependence on $u$:}
In Fig.~\ref{Sigma-Lambda-vs-uKA}c,d we show the variation of $\sigma^*$ and $\lambda^*$, 
respectively, as a function of $u$. 
We can see that $\sigma^*,\lambda^*$ decrease with $u$, as expected. 
It is notable that $\lambda^*$ converges to $1$ for every $u$ as the system size is increased.

\paragraph{Dependence on $K$:} 
Interesting results are obtained by studying the dependence on the number $K$ of
neighbors. Figure~\ref{Sigma-Lambda-vs-uKA}e reveals that $\sigma^*$ decreases 
with $K$, whereas Fig.~\ref{Sigma-Lambda-vs-uKA}f shows that $\lambda^*$ stays approximately 
constant. For increasing $K$, the system approaches a complete graph for which the annealed 
and quenched cases are equal. This means that correlations in the system decrease with 
increasing $K$ and $\sigma^*$ approaches $\lambda^*$. Again, $\lambda^* \rightarrow 1$ 
as $N$ grows.

\begin{figure}[!h]
\centering
\includegraphics[width=0.7\columnwidth]{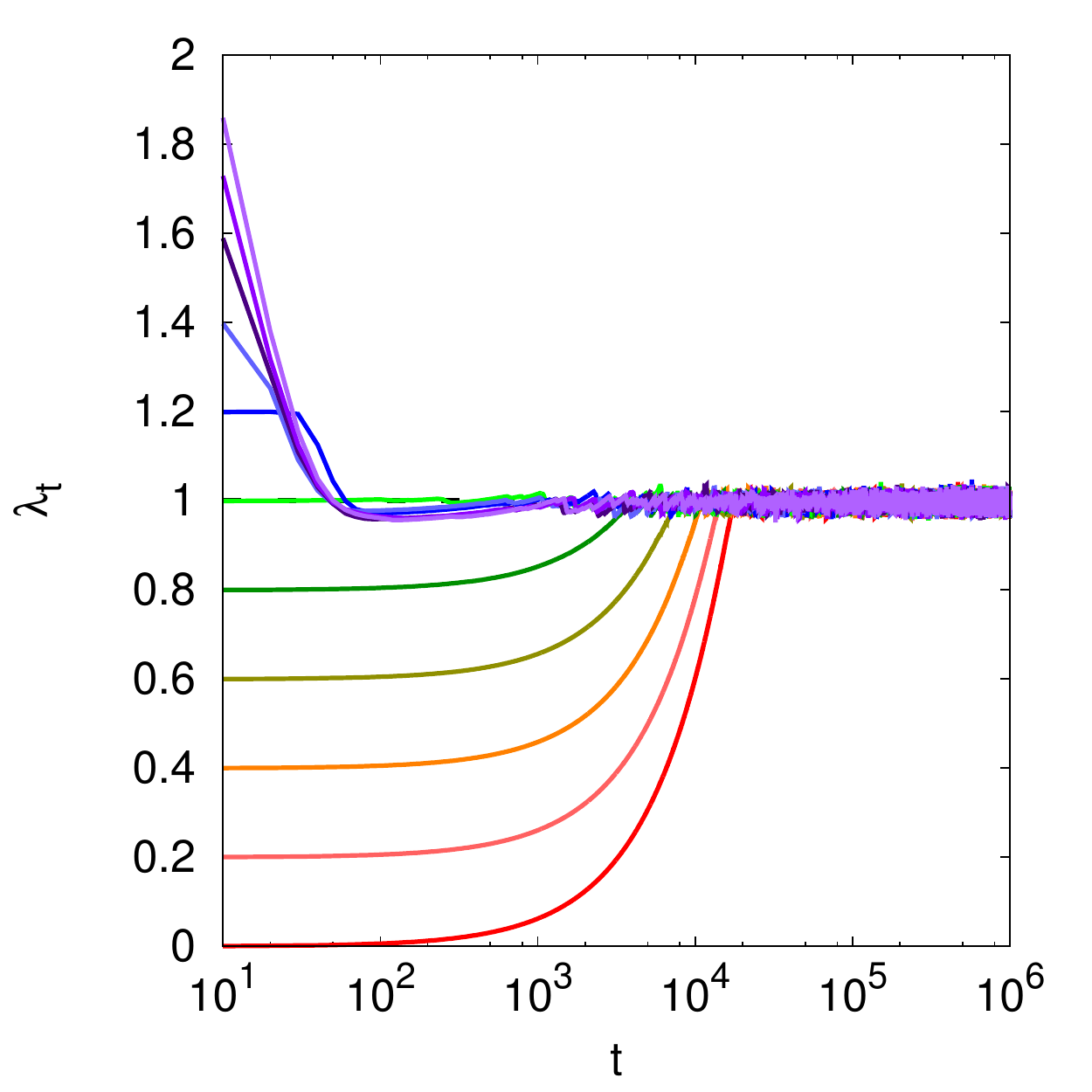}
\caption{Evolution of $\lambda_t$ versus $t$ for several different initial values $\lambda_0$. 
All curves reach a stationary mean value $\lambda^* \simeq 1$ after a transient time. 
Parameters: $N=32000$, $\epsilon=2$, $n=3$, $K=10$, $A=1.0$, $u=0.1$ and $a=1$.}
\label{lambdat}
\end{figure}

\subsection{The time series $\lambda_t$ and its fluctuations}
So, it seems that the stationary average value $\lambda^* = \left\langle \lambda_t \right\rangle_t$ 
achieves criticality for a  semi-infinite volume of the parameter 
space $(A,\epsilon,u,K)$. Hence, we focus on the behavior of the time series $\lambda_t$. 
In Fig.~\ref{lambdat} we plot $\lambda_t$ versus $t$, starting from different initial conditions 
$\lambda_0$. We see that, irrespective of the initial conditions, the network self-organizes  in a fast way
towards the critical value $\lambda^* \approx \lambda_c=1$.

\begin{figure}[!h]
\centering
\includegraphics[width=1.0\columnwidth]{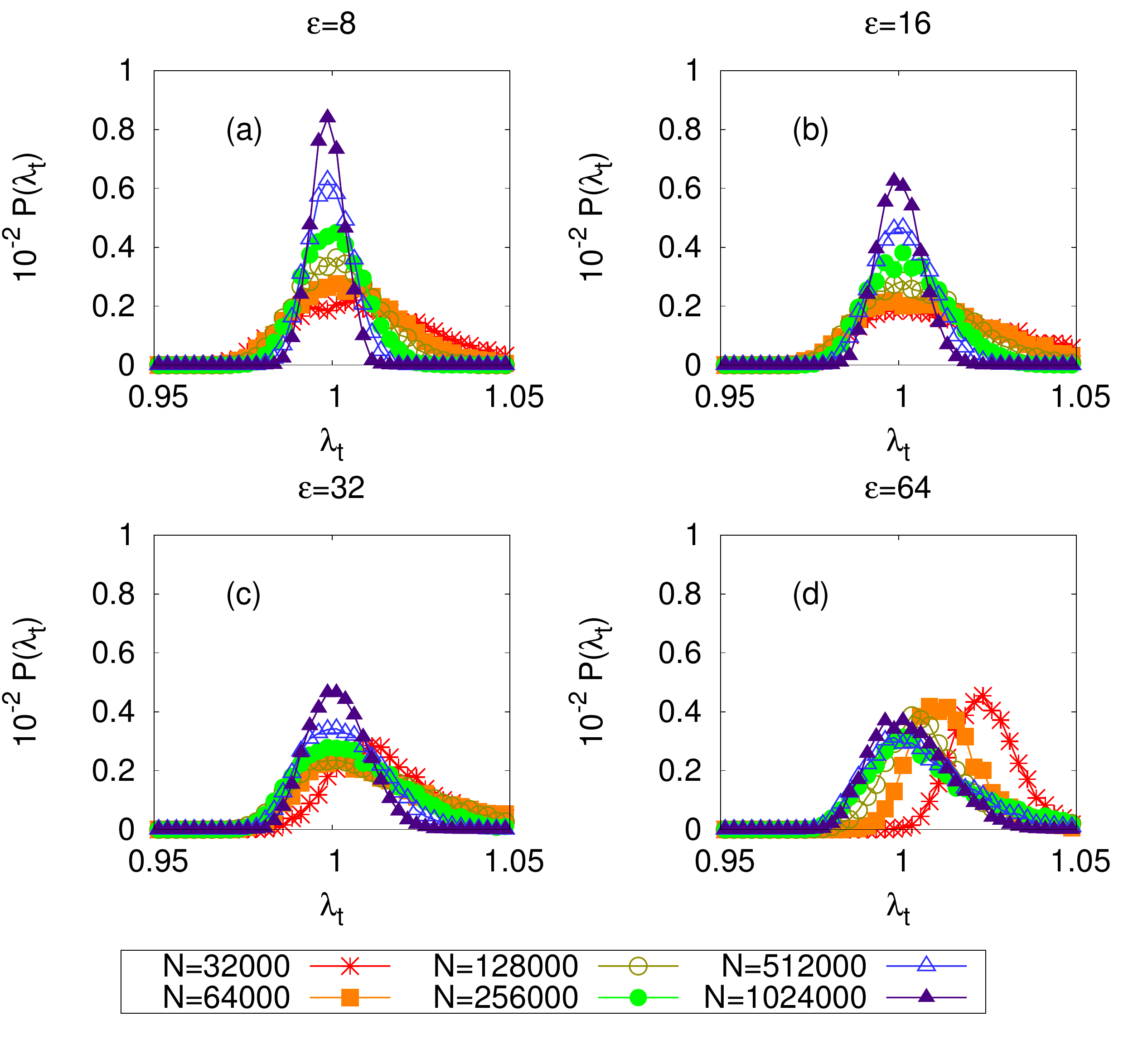}
\caption{Histograms of $\lambda_t$ for networks with quenched dynamics with fixed $\epsilon = 8, 16, 32$ and $64$. 
The lines are guides to the eye. Parameters: $n=3$, $K=10$, $A=1.0$, $u=0.1$ and $a=1$.}
\label{fig7}
\end{figure}

In Fig.~\ref{fig7} we present histograms for $\lambda_t$, in the stationary regime, for different 
values of $\epsilon$ and $N$. 
The data is collected every $100$ time steps during a time span of 
$10^6$ time steps, after a transient. 
We see that the width of the histograms decreases as $N$ grows. 
So we have a possible SOC behavior, similar to conservative models 
with static links~\cite{Bonachela09}. 
However, for $\epsilon=64$ this is not so apparent because $N$ is not sufficiently large (Fig.~\ref{fig7}d). 
This reveals that, although the model is very robust, for small networks the system is considerably dependent on the parameter space. 
In particular, the average $\lambda^*$ moves from supercritical values for small $N$ 
toward the critical value for large $N$.

\subsection{Scaling with system size}
\paragraph{The $a=2/3$ case:}
Bonachela \emph{et al.}~\cite{Bonachela10} found similarly strong finite size effects for the LHG model. 
They also found, both from a field theory and from simulations, that there exists a better 
scaling with $N$ which puts the network always at criticality, with robust power law avalanche size distributions, for any $N$. This occurs
if we use a $N^a$ scaling with exponent $a=2/3$, or equivalently, if we use $a=1$ 
with $\epsilon \propto N^{1/3}$~\cite{Bonachela10}.

\begin{figure}[!h]
\centering
\includegraphics[width=0.7\columnwidth]{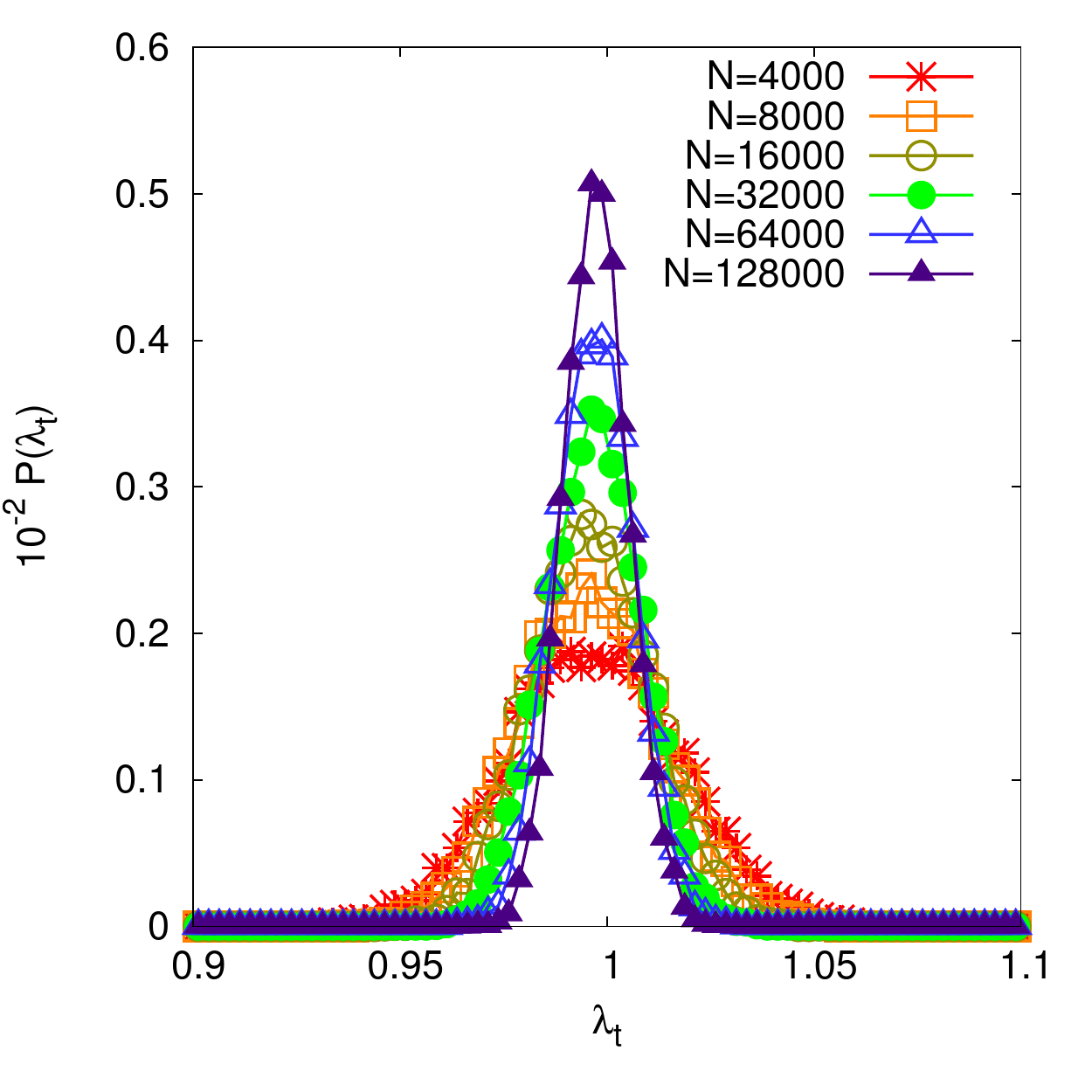}
\caption{Histograms of $\lambda_t$ for networks with quenched dynamics with $\epsilon=0.07 N^{1/3}$. 
The lines are guides to the eye. Parameters: $n=3$, $K=10$, $A=1.0$, $u=0.1$.}
\label{fig8}
\end{figure}

To check if this is also true for our model, we present similar histograms for $\lambda_t$ (Fig.~\ref{fig8}), but now using
the scaling $\epsilon = 0.07 N^{1/3}$ (or, equivalently, $\epsilon = 0.07, a = 2/3$). 
Indeed, with this scaling we obtain well behaved networks that are always critical, 
that is, have the average $\lambda^*$ equal to $1$ for all $N$. 
This peculiar exponent $1/3$ also appears in other models~\cite{Pruessner02,Bonachela09}, with similar results. 
In Fig.~\ref{fig8}, we see that $P(\lambda_t)$ sharpens as $N$ is increased.

Therefore, we conclude that, for $a>0$, the fluctuations of $\lambda_t$ around the mean 
value decrease for increasing system size and presumably vanish in the infinite-size limit. 
This result strongly differs from what is found for the LHG model~\cite{Bonachela10}, 
where fluctuations do not vanish for large $N$.
Indeed, due to this fact, Bonachela \emph{et al.} proposed that the LHG model pertains 
to the Dynamical Percolation universality class,
not the Directed Percolation class as standard SOC models. Our model, in contrast, 
clearly pertains to the Directed Percolation class, as shown by our mean field results,
but it is not clear what is the decisive difference between ours and the LHG model that produces 
such change of universality class. 
In fact, it would be interesting to extend this analysis to other analytically treatable models 
in which signs of criticality have been found, such as the non-conservative neuronal networks 
exhibiting up and down states proposed by Millman et al.~\cite{Millman10}. 
This is an open problem to be addressed in the future.

\paragraph*{The $a=0$ case:}
Finally, let us discuss the case $a=0$, which means that the synaptic recovery 
dynamics does not depends on $N$. Indeed, this is
the biologically realistic case since the recovery time cannot depend on 
non-local information like the network size $N$. Our mean-field result predicts:
\begin{equation}
    \sigma^* \simeq   1+ \frac{(AK-1)}{1+x},
\end{equation}
where now $x \equiv uK/[(n-1)\epsilon]$. Since $x$ now is always finite, the stationary 
value $\sigma^*$ is always supercritical (and we expect the same for $\lambda^*$). 
However, by using $AK \geq 1$, say $A= 1.1/K$, and a biologically 
motivated value for the number of synapses
($K = 10^4$), we obtain, for $n=3, u=1, \epsilon =4$:
\begin{equation}
    \sigma^* \simeq   1.0088 \:,
\end{equation}
which, although supercritical, seems to be sufficiently close to criticality to 
explain the experimental power laws in neuronal avalanches. Notice that this mean-field
result is also relevant to the quenched case since, as can be seen in 
Fig.~\ref{Sigma-Lambda-vs-uKA}e, 
we have $\sigma^*  \rightarrow \lambda_c = 1$ for large $K$.

This slight supercriticality has been called \emph{self-organized supercriticality} 
(SOSC) by Brochini \emph{et al.}~\cite{Brochini16}.
Curiously, superavalanches (the so called \emph{dragon kings}) and supercriticality 
have also been observed in experiments~\cite{deArcangelis12}.
This SOSC scenario is new and merits a proper study, with intensive simulations
as a function of the parametric space $(A,\epsilon,u )$, not done here.
This will be the subject of future work.


\section{Conclusion}

The general criticality condition is $\lambda_c = 1$, not $\sigma_c = 1$.
The branching ratio $\sigma$ is a good predictor of criticality only for annealed synaptic dynamics, 
where correlations are destroyed by construction. But there is a relation between $\lambda$ and $\sigma$
(for non-assortative networks, see~\cite{Larremore11a}): $\lambda = \eta \sigma$ where the correlation
coefficient $\eta$ can be larger or smaller than one, depending on the kind of correlations between the 
in-links and the out-links of the nodes. 

In our case, we found anti-correlation (and thus $\eta < 1, \lambda < \sigma$),
due to the fact that the avalanche dynamics induce that neurons with large sum of in-links fires more
and so their out-links are more depressed (and vice-versa). Of course, this scenario is not static,
the $(\sigma_i^{in},\sigma_i^{out})$ values vary with time, one can be larger than the other at some time
and the converse can also be true at another time. 
It is the ensemble average over this whole process that gives the final value of $\eta$.  

Notice that our network with dynamic links (and all other networks of the same kind~\cite{Arcangelis07,Levina07,Bonachela10,Costa15}) have two different aspects, self-organization
and criticality,  which are independent but sometimes confused in the literature:
\begin{itemize}
\item Self-organization: this is simply a word that describes the transient evolution of the synaptic distribution $P_t(P_{ij})$ 
from an initial condition $P_0(P_{ij})$ toward a stationary distribution $P^*(P_{ij})$. This stationary
distribution gives the final value for $\lambda^*$ that is not necessarly the critical one; 
\item Criticality: determination of some parametric volume that, at least in the $N \rightarrow \infty$ limit, gives 
$\lambda^* = \lambda_ c = 1$. For a \emph{bona fide} definition of SOC, this volume cannot be 
of zero measure, that is, we cannot have fine tuning.
Also, the fluctuations around $\lambda^*$ must vanish for large
networks (that is, the network sits at, not hovers around, the critical point). 
\end{itemize}

Concerning self-organization, the proposed mechanism of synaptic depression, Eq.~(\ref{Pij}), 
consistently points to a convergence to a 
self-organized regime $\lambda(t) \simeq \lambda^*$, both  for annealed as well 
as for quenched dynamics. 
This self-organized value $\lambda^*$ depends on the parameters 
$(A,\epsilon,u)$ and network size $N$, and is not necessarily critical. 

About criticality, the mean-field calculation suggests that there is a 
semi-infinite volume in the parameter space ($A,\epsilon,u$) that produces SOC 
behavior in the infinite size limit. This can be viewed in Eq.~(\ref{Omega-N}), 
where the dependence on all parameters ($A,\epsilon, u, n, K$)
vanishes for large $N$ (if $a>0$). This parametric volume is semi-infinite because low values of $\epsilon$
produce subcritical networks where Eq.~(\ref{Omega-N}) is no more valid.
So, the parametric space has two separated volumes: one subcritical and the other critical (SOC). 
Both are large and generic. In the thermodynamic limit, there are no parameters that produce supercritical activity.  

Intensive numerical simulations also lead us to the conclusion that, if we use an exponent $a >0$, 
then  there is a large volume in parameter space $(A,\epsilon,u)$ where 
$\left\langle \lambda(t) \right\rangle = \lambda^* = \lambda_c = 1$
when $N \rightarrow \infty$. Moreover, the fluctuations of $\lambda(t)$ around $\lambda^*$, as measured e.g. by the standard deviation of $P(\lambda(t))$, goes to zero as $N$ grows (an important property not found in the LHG model~\cite{Bonachela10}). 
So, both mean-field and simulations strongly suggest a
well behaved SOC scenario, at least in the thermodynamic limit,
with the presence of a Directed Percolation absorbing phase transition like other standard SOC models.

Finally, the case with $a=0$ only produces Self-Organized Supercriticality (SOSC~\cite{Brochini16}). 
However, for large number of synapses $K$, as suggested by biology, networks which are 
almost critical are obtained, and this can be sufficient to deal with the power 
laws found in experiments.
Moreover, the SOSC scenario suggests that biological neuronal networks could be indeed 
slightly supercritical, a fact perhaps masked by 
standard experiments with few electrodes but sometimes revealed in dragon king 
avalanches~\cite{deArcangelis12}. This SOSC scenario (exponent $a=0$ case) should be 
studied more deeply in another publication.

\begin{acknowledgments}
This article was produced as part of the activities of FAPESP  Research, Innovation 
and Dissemination Center for Neuromathematics (grant \#2013/07699-0, S.Paulo Research Foundation). 
We acknowledge financial support from CAPES, CNPq, FACEPE, and Centre for Natural and Artifical Information
Processing Systems (CNAIPS)-USP. AAC thanks FAPESP (grant \#2016/00430-3 and \#2016/20945-8).

\end{acknowledgments}



\begin{thebibliography}{24}%
\makeatletter
\providecommand \@ifxundefined [1]{%
 \@ifx{#1\undefined}
}%
\providecommand \@ifnum [1]{%
 \ifnum #1\expandafter \@firstoftwo
 \else \expandafter \@secondoftwo
 \fi
}%
\providecommand \@ifx [1]{%
 \ifx #1\expandafter \@firstoftwo
 \else \expandafter \@secondoftwo
 \fi
}%
\providecommand \natexlab [1]{#1}%
\providecommand \enquote  [1]{``#1''}%
\providecommand \bibnamefont  [1]{#1}%
\providecommand \bibfnamefont [1]{#1}%
\providecommand \citenamefont [1]{#1}%
\providecommand \href@noop [0]{\@secondoftwo}%
\providecommand \href [0]{\begingroup \@sanitize@url \@href}%
\providecommand \@href[1]{\@@startlink{#1}\@@href}%
\providecommand \@@href[1]{\endgroup#1\@@endlink}%
\providecommand \@sanitize@url [0]{\catcode `\\12\catcode `\$12\catcode
  `\&12\catcode `\#12\catcode `\^12\catcode `\_12\catcode `\%12\relax}%
\providecommand \@@startlink[1]{}%
\providecommand \@@endlink[0]{}%
\providecommand \url  [0]{\begingroup\@sanitize@url \@url }%
\providecommand \@url [1]{\endgroup\@href {#1}{\urlprefix }}%
\providecommand \urlprefix  [0]{URL }%
\providecommand \Eprint [0]{\href }%
\providecommand \doibase [0]{http://dx.doi.org/}%
\providecommand \selectlanguage [0]{\@gobble}%
\providecommand \bibinfo  [0]{\@secondoftwo}%
\providecommand \bibfield  [0]{\@secondoftwo}%
\providecommand \translation [1]{[#1]}%
\providecommand \BibitemOpen [0]{}%
\providecommand \bibitemStop [0]{}%
\providecommand \bibitemNoStop [0]{.\EOS\space}%
\providecommand \EOS [0]{\spacefactor3000\relax}%
\providecommand \BibitemShut  [1]{\csname bibitem#1\endcsname}%
\let\auto@bib@innerbib\@empty
\bibitem [{\citenamefont {Beggs}\ and\ \citenamefont {Plenz}(2003)}]{Beggs03}%
  \BibitemOpen
  \bibfield  {author} {\bibinfo {author} {\bibfnamefont {J.~M.}\ \bibnamefont
  {Beggs}}\ and\ \bibinfo {author} {\bibfnamefont {D.}~\bibnamefont {Plenz}},\
  }\href@noop {} {\bibfield  {journal} {\bibinfo  {journal} {J. Neurosci.}\
  }\textbf {\bibinfo {volume} {23}},\ \bibinfo {pages} {11167} (\bibinfo {year}
  {2003})}\BibitemShut {NoStop}%
\bibitem [{\citenamefont {Bertschinger}\ and\ \citenamefont
  {Natschl\"ager}(2004)}]{Bertschinger04}%
  \BibitemOpen
  \bibfield  {author} {\bibinfo {author} {\bibfnamefont {N.}~\bibnamefont
  {Bertschinger}}\ and\ \bibinfo {author} {\bibfnamefont {T.}~\bibnamefont
  {Natschl\"ager}},\ }\href@noop {} {\bibfield  {journal} {\bibinfo  {journal}
  {Neural Comput.}\ }\textbf {\bibinfo {volume} {16}},\ \bibinfo {pages} {1413}
  (\bibinfo {year} {2004})}\BibitemShut {NoStop}%
\bibitem [{\citenamefont {Shew}\ \emph {et~al.}(2011)\citenamefont {Shew},
  \citenamefont {Yang}, \citenamefont {Yu}, \citenamefont {Roy},\ and\
  \citenamefont {Plenz}}]{Shew11}%
  \BibitemOpen
  \bibfield  {author} {\bibinfo {author} {\bibfnamefont {W.~L.}\ \bibnamefont
  {Shew}}, \bibinfo {author} {\bibfnamefont {H.}~\bibnamefont {Yang}}, \bibinfo
  {author} {\bibfnamefont {S.}~\bibnamefont {Yu}}, \bibinfo {author}
  {\bibfnamefont {R.}~\bibnamefont {Roy}}, \ and\ \bibinfo {author}
  {\bibfnamefont {D.}~\bibnamefont {Plenz}},\ }\href@noop {} {\bibfield
  {journal} {\bibinfo  {journal} {J. Neurosci.}\ }\textbf {\bibinfo {volume}
  {31}},\ \bibinfo {pages} {55} (\bibinfo {year} {2011})}\BibitemShut {NoStop}%
\bibitem [{\citenamefont {Kinouchi}\ and\ \citenamefont
  {Copelli}(2006)}]{Kinouchi06a}%
  \BibitemOpen
  \bibfield  {author} {\bibinfo {author} {\bibfnamefont {O.}~\bibnamefont
  {Kinouchi}}\ and\ \bibinfo {author} {\bibfnamefont {M.}~\bibnamefont
  {Copelli}},\ }\href {\doibase 10.1038/nphys289} {\bibfield  {journal}
  {\bibinfo  {journal} {Nat. Phys.}\ }\textbf {\bibinfo {volume} {2}},\
  \bibinfo {pages} {348} (\bibinfo {year} {2006})}\BibitemShut {NoStop}%
\bibitem [{\citenamefont {Shew}\ \emph {et~al.}(2009)\citenamefont {Shew},
  \citenamefont {Yang}, \citenamefont {Petermann}, \citenamefont {Roy},\ and\
  \citenamefont {Plenz}}]{Shew09}%
  \BibitemOpen
  \bibfield  {author} {\bibinfo {author} {\bibfnamefont {W.}~\bibnamefont
  {Shew}}, \bibinfo {author} {\bibfnamefont {H.}~\bibnamefont {Yang}}, \bibinfo
  {author} {\bibfnamefont {T.}~\bibnamefont {Petermann}}, \bibinfo {author}
  {\bibfnamefont {R.}~\bibnamefont {Roy}}, \ and\ \bibinfo {author}
  {\bibfnamefont {D.}~\bibnamefont {Plenz}},\ }\href {\doibase
  10.1523/JNEUROSCI.3864-09.2009} {\bibfield  {journal} {\bibinfo  {journal}
  {J. Neurosci.}\ }\textbf {\bibinfo {volume} {29}},\ \bibinfo {pages} {15595}
  (\bibinfo {year} {2009})}\BibitemShut {NoStop}%
\bibitem [{\citenamefont {Gautam}\ \emph {et~al.}(2015)\citenamefont {Gautam},
  \citenamefont {Hoang}, \citenamefont {McClanahan}, \citenamefont {Grady},\
  and\ \citenamefont {Shew}}]{Gautam2015}%
  \BibitemOpen
  \bibfield  {author} {\bibinfo {author} {\bibfnamefont {S.~H.}\ \bibnamefont
  {Gautam}}, \bibinfo {author} {\bibfnamefont {T.~T.}\ \bibnamefont {Hoang}},
  \bibinfo {author} {\bibfnamefont {K.}~\bibnamefont {McClanahan}}, \bibinfo
  {author} {\bibfnamefont {S.~K.}\ \bibnamefont {Grady}}, \ and\ \bibinfo
  {author} {\bibfnamefont {W.~L.}\ \bibnamefont {Shew}},\ }\href {\doibase
  10.1371/journal.pcbi.1004576} {\bibfield  {journal} {\bibinfo  {journal}
  {PLoS Comput Biol}\ }\textbf {\bibinfo {volume} {11}},\ \bibinfo {pages} {1}
  (\bibinfo {year} {2015})}\BibitemShut {NoStop}%
\bibitem [{\citenamefont {Girardi-Schappo}\ \emph {et~al.}(2016)\citenamefont
  {Girardi-Schappo}, \citenamefont {Bortolotto}, \citenamefont {Gonsalves},
  \citenamefont {Pinto},\ and\ \citenamefont
  {Tragtenberg}}]{GirardiShappo2016}%
  \BibitemOpen
  \bibfield  {author} {\bibinfo {author} {\bibfnamefont {M.}~\bibnamefont
  {Girardi-Schappo}}, \bibinfo {author} {\bibfnamefont {G.~S.}\ \bibnamefont
  {Bortolotto}}, \bibinfo {author} {\bibfnamefont {J.~J.}\ \bibnamefont
  {Gonsalves}}, \bibinfo {author} {\bibfnamefont {L.~T.}\ \bibnamefont
  {Pinto}}, \ and\ \bibinfo {author} {\bibfnamefont {M.~H.~R.}\ \bibnamefont
  {Tragtenberg}},\ }\href@noop {} {\bibfield  {journal} {\bibinfo  {journal}
  {Sci. Rep.}\ }\textbf {\bibinfo {volume} {6}} (\bibinfo {year}
  {2016})}\BibitemShut {NoStop}%
\bibitem [{\citenamefont {Chialvo}(2010)}]{Chialvo10}%
  \BibitemOpen
  \bibfield  {author} {\bibinfo {author} {\bibfnamefont {D.~R.}\ \bibnamefont
  {Chialvo}},\ }\href@noop {} {\bibfield  {journal} {\bibinfo  {journal} {Nat.
  Phys.}\ }\textbf {\bibinfo {volume} {6}},\ \bibinfo {pages} {744} (\bibinfo
  {year} {2010})}\BibitemShut {NoStop}%
\bibitem [{\citenamefont {Shew}\ and\ \citenamefont {Plenz}(2013)}]{Shew13}%
  \BibitemOpen
  \bibfield  {author} {\bibinfo {author} {\bibfnamefont {W.}~\bibnamefont
  {Shew}}\ and\ \bibinfo {author} {\bibfnamefont {D.}~\bibnamefont {Plenz}},\
  }\href {\doibase 10.1177/1073858412445487} {\bibfield  {journal} {\bibinfo
  {journal} {Neuroscientist}\ }\textbf {\bibinfo {volume} {19}},\ \bibinfo
  {pages} {88} (\bibinfo {year} {2013})}\BibitemShut {NoStop}%
\bibitem [{\citenamefont {de~Arcangelis}\ \emph {et~al.}(2006)\citenamefont
  {de~Arcangelis}, \citenamefont {Perrone-Capano},\ and\ \citenamefont
  {Herrmann}}]{deArcangelis06}%
  \BibitemOpen
  \bibfield  {author} {\bibinfo {author} {\bibfnamefont {L.}~\bibnamefont
  {de~Arcangelis}}, \bibinfo {author} {\bibfnamefont {C.}~\bibnamefont
  {Perrone-Capano}}, \ and\ \bibinfo {author} {\bibfnamefont {H.~J.}\
  \bibnamefont {Herrmann}},\ }\href {\doibase 10.1103/PhysRevLett.96.028107}
  {\bibfield  {journal} {\bibinfo  {journal} {Phys. Rev. Lett.}\ }\textbf
  {\bibinfo {volume} {96}},\ \bibinfo {pages} {028107} (\bibinfo {year}
  {2006})}\BibitemShut {NoStop}%
\bibitem [{\citenamefont {de~Arcangelis}(2012)}]{deArcangelis12}%
  \BibitemOpen
  \bibfield  {author} {\bibinfo {author} {\bibfnamefont {L.}~\bibnamefont
  {de~Arcangelis}},\ }\href@noop {} {\bibfield  {journal} {\bibinfo  {journal}
  {Eur. Phys. J. Spec. Top.}\ }\textbf {\bibinfo {volume} {205}},\ \bibinfo
  {pages} {243} (\bibinfo {year} {2012})}\BibitemShut {NoStop}%
\bibitem [{\citenamefont {Lombardi}\ \emph {et~al.}(2016)\citenamefont
  {Lombardi}, \citenamefont {Herrmann}, \citenamefont {Plenz},\ and\
  \citenamefont {de~Arcangelis}}]{deArcangelis16}%
  \BibitemOpen
  \bibfield  {author} {\bibinfo {author} {\bibfnamefont {F.}~\bibnamefont
  {Lombardi}}, \bibinfo {author} {\bibfnamefont {H.~J.}\ \bibnamefont
  {Herrmann}}, \bibinfo {author} {\bibfnamefont {D.}~\bibnamefont {Plenz}}, \
  and\ \bibinfo {author} {\bibfnamefont {L.}~\bibnamefont {de~Arcangelis}},\
  }\href@noop {} {\bibfield  {journal} {\bibinfo  {journal} {Sci. Rep.}\
  }\textbf {\bibinfo {volume} {6}},\ \bibinfo {pages} {24690} (\bibinfo {year}
  {2016})}\BibitemShut {NoStop}%
\bibitem [{\citenamefont {Levina}\ \emph {et~al.}(2007)\citenamefont {Levina},
  \citenamefont {Herrmann},\ and\ \citenamefont {Geisel}}]{Levina07}%
  \BibitemOpen
  \bibfield  {author} {\bibinfo {author} {\bibfnamefont {A.}~\bibnamefont
  {Levina}}, \bibinfo {author} {\bibfnamefont {J.~M.}\ \bibnamefont
  {Herrmann}}, \ and\ \bibinfo {author} {\bibfnamefont {T.}~\bibnamefont
  {Geisel}},\ }\href {\doibase 10.1038/nphys:758} {\bibfield  {journal}
  {\bibinfo  {journal} {Nat. Phys.}\ }\textbf {\bibinfo {volume} {3}},\
  \bibinfo {pages} {857} (\bibinfo {year} {2007})}\BibitemShut {NoStop}%
\bibitem [{\citenamefont {Bonachela}\ and\ \citenamefont
  {Mu\~noz}(2009)}]{Bonachela09}%
  \BibitemOpen
  \bibfield  {author} {\bibinfo {author} {\bibfnamefont {J.~A.}\ \bibnamefont
  {Bonachela}}\ and\ \bibinfo {author} {\bibfnamefont {M.~A.}\ \bibnamefont
  {Mu\~noz}},\ }\href@noop {} {\bibfield  {journal} {\bibinfo  {journal} {J.
  Stat. Mech.}\ }\textbf {\bibinfo {volume} {2009}},\ \bibinfo {pages} {P09009}
  (\bibinfo {year} {2009})}\BibitemShut {NoStop}%
\bibitem [{\citenamefont {Bonachela}\ \emph {et~al.}(2010)\citenamefont
  {Bonachela}, \citenamefont {de~Franciscis}, \citenamefont {Torres},\ and\
  \citenamefont {Mu\~noz}}]{Bonachela10}%
  \BibitemOpen
  \bibfield  {author} {\bibinfo {author} {\bibfnamefont {J.~A.}\ \bibnamefont
  {Bonachela}}, \bibinfo {author} {\bibfnamefont {S.}~\bibnamefont
  {de~Franciscis}}, \bibinfo {author} {\bibfnamefont {J.~J.}\ \bibnamefont
  {Torres}}, \ and\ \bibinfo {author} {\bibfnamefont {M.~A.}\ \bibnamefont
  {Mu\~noz}},\ }\href@noop {} {\bibfield  {journal} {\bibinfo  {journal} {J.
  Stat. Mech.}\ }\textbf {\bibinfo {volume} {2010}},\ \bibinfo {pages} {P02015}
  (\bibinfo {year} {2010})}\BibitemShut {NoStop}%
\bibitem [{\citenamefont {Moosavi}\ and\ \citenamefont
  {Montakhab}(2014)}]{Moosavi&Montakhab}%
  \BibitemOpen
  \bibfield  {author} {\bibinfo {author} {\bibfnamefont {S.~A.}\ \bibnamefont
  {Moosavi}}\ and\ \bibinfo {author} {\bibfnamefont {A.}~\bibnamefont
  {Montakhab}},\ }\href {\doibase 10.1103/PhysRevE.89.052139} {\bibfield
  {journal} {\bibinfo  {journal} {Phys. Rev. E}\ }\textbf {\bibinfo {volume}
  {89}},\ \bibinfo {pages} {052139} (\bibinfo {year} {2014})}\BibitemShut
  {NoStop}%
\bibitem [{\citenamefont {Costa}\ \emph {et~al.}(2015)\citenamefont {Costa},
  \citenamefont {Copelli},\ and\ \citenamefont {Kinouchi}}]{Costa15}%
  \BibitemOpen
  \bibfield  {author} {\bibinfo {author} {\bibfnamefont {A.~A.}\ \bibnamefont
  {Costa}}, \bibinfo {author} {\bibfnamefont {M.}~\bibnamefont {Copelli}}, \
  and\ \bibinfo {author} {\bibfnamefont {O.}~\bibnamefont {Kinouchi}},\
  }\href@noop {} {\bibfield  {journal} {\bibinfo  {journal} {J. Stat. Mech.}\
  }\textbf {\bibinfo {volume} {2015}},\ \bibinfo {pages} {P06004} (\bibinfo
  {year} {2015})}\BibitemShut {NoStop}%
\bibitem [{\citenamefont {Larremore}\ \emph {et~al.}(2011)\citenamefont
  {Larremore}, \citenamefont {Shew},\ and\ \citenamefont
  {Restrepo}}]{Larremore11a}%
  \BibitemOpen
  \bibfield  {author} {\bibinfo {author} {\bibfnamefont {D. B.}~\bibnamefont
  {Larremore}}, \bibinfo {author} {\bibfnamefont {W. L.}~\bibnamefont {Shew}}, \
  and\ \bibinfo {author} {\bibfnamefont {J. G.}~\bibnamefont {Restrepo}},\ }\href
  {\doibase 10.1103/PhysRevLett.106.058101} {\bibfield  {journal} {\bibinfo
  {journal} {Phys. Rev. Lett.}\ }\textbf {\bibinfo {volume} {106}},\ \bibinfo
  {pages} {058101} (\bibinfo {year} {2011})}\BibitemShut {NoStop}%
\bibitem [{\citenamefont {Restrepo}\ \emph {et~al.}(2007)\citenamefont
  {Restrepo}, \citenamefont {Ott},\ and\ \citenamefont {Hunt}}]{Restrepo07}%
  \BibitemOpen
  \bibfield  {author} {\bibinfo {author} {\bibfnamefont {J.~G.}\ \bibnamefont
  {Restrepo}}, \bibinfo {author} {\bibfnamefont {E.}~\bibnamefont {Ott}}, \
  and\ \bibinfo {author} {\bibfnamefont {B.~R.}\ \bibnamefont {Hunt}},\
  }\href@noop {} {\bibfield  {journal} {\bibinfo  {journal} {Phys. Rev. E}\
  }\textbf {\bibinfo {volume} {76}},\ \bibinfo {pages} {056119} (\bibinfo
  {year} {2007})}\BibitemShut {NoStop}%
\bibitem [{\citenamefont {Harris}(2002)}]{Harris02}%
  \BibitemOpen
  \bibfield  {author} {\bibinfo {author} {\bibfnamefont {T.~E.}\ \bibnamefont
  {Harris}},\ }\href@noop {} {\emph {\bibinfo {title} {The theory of branching
  processes}}}\ (\bibinfo  {publisher} {Courier Corporation},\ \bibinfo {year}
  {2002})\BibitemShut {NoStop}%
\bibitem [{\citenamefont {Pruessner}\ and\ \citenamefont
  {Jensen}(2002)}]{Pruessner02}%
  \BibitemOpen
  \bibfield  {author} {\bibinfo {author} {\bibfnamefont {G.}~\bibnamefont
  {Pruessner}}\ and\ \bibinfo {author} {\bibfnamefont {H.~J.}\ \bibnamefont
  {Jensen}},\ }\href@noop {} {\bibfield  {journal} {\bibinfo  {journal} {EPL
  (Europhysics Letters)}\ }\textbf {\bibinfo {volume} {58}},\ \bibinfo {pages}
  {250} (\bibinfo {year} {2002})}\BibitemShut {NoStop}%
\bibitem [{\citenamefont {Millman}\ \emph {et~al.}(2010)\citenamefont
  {Millman}, \citenamefont {Mihalas}, \citenamefont {Kirkwood},\ and\
  \citenamefont {Niebur}}]{Millman10}%
  \BibitemOpen
  \bibfield  {author} {\bibinfo {author} {\bibfnamefont {D.}~\bibnamefont
  {Millman}}, \bibinfo {author} {\bibfnamefont {S.}~\bibnamefont {Mihalas}},
  \bibinfo {author} {\bibfnamefont {A.}~\bibnamefont {Kirkwood}}, \ and\
  \bibinfo {author} {\bibfnamefont {E.}~\bibnamefont {Niebur}},\ }\href
  {\doibase 10.1038/nphys1757} {\bibfield  {journal} {\bibinfo  {journal} {Nat.
  Phys.}\ }\textbf {\bibinfo {volume} {6}},\ \bibinfo {pages} {801} (\bibinfo
  {year} {2010})}\BibitemShut {NoStop}%
\bibitem [{\citenamefont {Brochini}\ \emph {et~al.}(2016)\citenamefont
  {Brochini}, \citenamefont {Costa}, \citenamefont {Abadi}, \citenamefont
  {Roque}, \citenamefont {Stolfi},\ and\ \citenamefont
  {Kinouchi}}]{Brochini16}%
  \BibitemOpen
  \bibfield  {author} {\bibinfo {author} {\bibfnamefont {L.}~\bibnamefont
  {Brochini}}, \bibinfo {author} {\bibfnamefont {A.~A.}\ \bibnamefont {Costa}},
  \bibinfo {author} {\bibfnamefont {M.}~\bibnamefont {Abadi}}, \bibinfo
  {author} {\bibfnamefont {A.~C.}\ \bibnamefont {Roque}}, \bibinfo {author}
  {\bibfnamefont {J.}~\bibnamefont {Stolfi}}, \ and\ \bibinfo {author}
  {\bibfnamefont {O.}~\bibnamefont {Kinouchi}},\ }\href@noop {} {\bibfield
  {journal} {\bibinfo  {journal} {Sci. Rep.}\ }\textbf {\bibinfo {volume}
  {6}},\ \bibinfo {pages} {35831} (\bibinfo {year} {2016})}\BibitemShut
  {NoStop}%
\bibitem [{\citenamefont {Pellegrini}\ \emph {et~al.}(2007)\citenamefont
  {Pellegrini}, \citenamefont {de~Arcangelis}, \citenamefont {Herrmann},\ and\
  \citenamefont {Perrone-Capano}}]{Arcangelis07}%
  \BibitemOpen
  \bibfield  {author} {\bibinfo {author} {\bibfnamefont {G.~L.}\ \bibnamefont
  {Pellegrini}}, \bibinfo {author} {\bibfnamefont {L.}~\bibnamefont
  {de~Arcangelis}}, \bibinfo {author} {\bibfnamefont {H.~J.}\ \bibnamefont
  {Herrmann}}, \ and\ \bibinfo {author} {\bibfnamefont {C.}~\bibnamefont
  {Perrone-Capano}},\ }\href {\doibase 10.1103/PhysRevE.76.016107} {\bibfield
  {journal} {\bibinfo  {journal} {Phys. Rev. E}\ }\textbf {\bibinfo {volume}
  {76}},\ \bibinfo {pages} {016107} (\bibinfo {year} {2007})}\BibitemShut
  {NoStop}%
\end{thebibliography}

%

\end{document}